# Role of momentum dependent interactions in nuclear stopping

*Mandeep Kaur[1], Sanjeev Kumar[2] Suneel Kumar[2]

[1]Shaheed Udham Singh College of Engineering & Technologyy, Tangori. Mohali 140306, Punjab, INDIA
[2] School of Physics and Materials Science, Thapar University, Patiala 147004, Punjab, INDIA
*email: mandeep062000@gmail.com

## Introduction

The study of heavy ion collision (HIC) at intermediate energies develops a great interest to understand the behavior of hot and dense nuclear matter under the extreme conditions of temperature, pressure and densities. In addition to this, it also helps us to study the time evolution of the reaction. During the last few years, studies were mainly focused on the origin of different mass fragments as well as the movement of particles along different directions.

Nuclear stopping is also one of the major phenomena happening at intermediate energy HIC. The study of nuclear stopping is one of the central topics of present scenario. It is the transformation of nuclear matter from a fused state to its disassembly. The complete destruction of the nuclear matter is termed as global stopping. One has also tried to correlate the light charged particles (LCP) with nuclear stopping parameters in HIC using QMD and IQMD models. Kumar et al. had tried to correlate it using simple static equation of state (EOS).[1]

Momentum dependent interactions (MDI) play a very important role in determining the nuclear dynamics. The initial attempts with MDI have showed a drastic effect on the collective flow as well as the particle production (i.e. it results in large number of fragments at peripheral collisions.). The effect of MDI is strongest during the initial phase of the reaction. The particles propagating with MDI are accelerated in the transverse direction during the early phase of the reaction. Our main motive is to study nuclear stopping taking into account the momentum dependence of the reaction.

## IQMD Model:

The Isospin-dependent Quantum Molecular Dynamic model is the refinement of QMD model based on event by event method. The reaction dynamics are governed by mean field, two-body collision and Pauli blocking.

**Initialization:** The baryons are represented by Gaussian-shaped density distributions

$$f_i(\vec{r},\vec{p},t) = \frac{1}{\pi^2 \hbar^2} e^{-[\vec{r}-\vec{r}_i(t)]^2 \frac{1}{2L}} e^{-[\vec{p}-\vec{p}_i(t)]^2 \frac{2L}{\hbar^2}}$$

**Propagation:** The successfully initialized nuclei are then boosted towards each other using Hamilton equations of motion

$$\frac{dr_i}{dt} = \frac{d\langle H \rangle}{d p_i} \ ; \ \frac{dp_i}{dt} = -\frac{d\langle H \rangle}{d r_i}$$

With $\langle H \rangle = \langle T \rangle + \langle V \rangle$ is the total Hamiltonian.

$$\langle H \rangle = \sum_i \frac{p_i^2}{2m_i} + \sum_i \sum_{j>i} \int f_i(\vec{r},\vec{p},t) V^{ij}(\vec{r'},\vec{r})$$

$$\times f_j(\vec{r'},\vec{p'},t) d\vec{r} d\vec{r'} d\vec{p} d\vec{p'}$$

The total potential is the sum of the following specific elementary potentials.

$$V = V_{Sky} + V_{Yuk} + V_{Coul} + V_{mdi} + V_{loc}$$

**Collision:** During the propagation, two nucleons are supposed to suffer a binary collision if the distance between their centroid is

$$|r_i - r_j| \leq \sqrt{\frac{\sigma_{tot}}{\pi}}$$

Where $\sigma_{tot} = \sigma(\sqrt{s}, type)$

The collision is blocked with a possibility
$$P_{block} = 1-(1-P_i)(1-P_j)$$

Where $P_i$ and $P_j$ are the already occupied phase space fractions by other nucleons.

**Result and discussion:**

The degree of stopping can be described with the help of different parameters which are discussed below.

**Anisotropic ratio:**
$$R = \frac{2}{\pi} \frac{[\sum_i^A |p_\perp(i)|]}{[\sum_i^A |p_\parallel(i)|]}$$
Where $p_\perp = \sqrt{p_x^2 + p_y^2}$ and $p_\parallel = p_z$

**Quadrupole moment:**
$$Q_{zz} = \sum_i^A [2p_z^2(i) - p_x^2(i) - p_y^2(i)]$$

**Rapidity distribution:**
$$Y(i) = \frac{1}{2} \ln \frac{E(i)+p_z(i)}{E(i)-p_z(i)}$$

Here, $E(i)$ and $p_z(i)$ are the energy and momentum of $i^{th}$ particle.

In the present analysis, thousand of events were simulated for the neutron rich reaction of $_{54}Xe^{131} + _{54}Xe^{131}$, at different impact parameters using hard (HMD) and soft (SMD) momentum dependent equation of state. The displayed result is at $E_{sym} = 32$ MeV.

In fig.1, We check the sensitivity of nuclear stopping as well as the multiplicity of free particles and LCP's for different nuclear equation of states with MDI's. For this purpose, a hard momentum dependent (HMD) and soft momentum dependent (SMD) equation of state with compressibility $\kappa = 380$ and $200$ MeV are employed, respectively. Repulsive interactions are stronger with HMD compared to SMD. Since at lower impact parameter the participant zone is more. Hence one can clearly see more stopping in the presence of HMD. This difference keeps on decreasing as one move from central to peripheral collisions. Production of LCP's and free particles follows the same trend as stopping parameter R. This simply implies that the fragment production can act as a global indicator for the nuclear stopping.

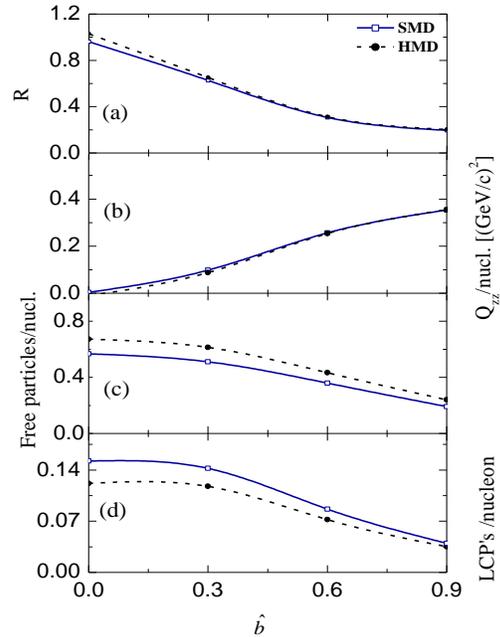

**Fig.1.** Impact parameter dependence of R (a), $\frac{1}{Q_{zz}/nucl.}$ (b), free nucleons(c) and LCP's (d) with hard (HMD) and soft (SMD) equation of state with symmetry energy=32 MeV.